# An Electric Vehicle Control Strategy to Mitigate Load Altering Attacks Against Power Grids

Mohammad Ali Sayed, Mohsen Ghafouri, Ribal Atallah, Mourad Debbabi, Chadi Assi

*Abstract*—Due to growing environmental concerns, the world's governments have been encouraging the shift of the transportation sector towards the adoption of Electric Vehicles (EVs). As a result, EV numbers have been growing exponentially and are expected to continue growing further which will add a large EV charging load to the power grid. To this end, this paper presents an EV-based defense mechanism against Load-Altering (LA) attacks targeting the grid. The developed mechanism utilizes H-∞ controllers and Linear Matrix Inequalities (LMIs) to mitigate LA attacks. After the controller synthesis and presentation of the attack scenarios, we demonstrate the effectiveness and success of our defense mechanism against the three known types of LA attacks. The scenarios include three 800 MW LA attacks against the New England 39-bus grid. The results demonstrate how our EV-based mitigation scheme eliminates the attack impacts and maintains the grid's stability in face of an unknown persisting attack.

## I. Introduction

HUMANITY's increasing reliance on electricity has placed the power grid at the center of any advanced society. To this end, smart technologies have been introduced to support grid operation transforming it into a smart grid [1]. However, this new interconnected grid has become vulnerable to attacks initiated through its new cyberinfrastructure.

The most common attack studied in the literature is the False Data Injection (FDI) attack [2] which tampers with the grid's measurements to alter the grid's state estimation and cause operators to take actions that might damage the grid. As a result, multiple attempts have been made to secure the communication layer of the power grid against such attacks [3]. Yet Load-Altering (LA) attacks target the grid's demand side rather than the state estimation and can only be seen through their impact [4], bypassing the grid's bad data detection mechanisms. LA attacks are classified into 3 broad subfamilies which are static attacks [4] [5], switching attacks [6], and dynamic attacks [7].

The authors of [4] demonstrated how static attacks, based on a single spike in load, can be initiated by manipulating high-wattage home IoT devices to cause line tripping in the grid while remaining unobservable to the utility. The work in [6] relied on the manipulation of distribution feeders to initiate switching attacks while mimicking natural phenomena on the grid which makes them hard to detect. These examples, however, should not overweigh the advantages of the smart grid and its incorporated smart technologies. One such technology is the Electric Vehicle (EV) and its charging infrastructure.

Faced with the susceptibility of the grid to LA attacks, we propose a mitigation scheme that takes advantage of the EVs' unique properties against such attacks. EVs can act as distributed batteries and generators that store energy and inject it into the grid using the Vehicle-to-Grid (V2G) power flow capability in new EV Charging Stations (EVCSs). EVs are spread throughout the power grid such that their distribution covers most if not all the load buses making them collocated with the attacker-controlled loads. This colocation allows the EVs to mitigate the attacks with minimal disturbance propagation through the grid. Finally, EVCSs have another edge over turbine generators whose reaction time is dictated by their weight, size, and control mechanisms. EVCSs are based on bidirectional power converters [8] that can change their charging rate and toggle between different states almost instantaneously. This makes EVs ideal for responding to LA attacks initiated using IoT loads based on similar converters.

In this paper, we create a wide area H-∞ controller that utilizes EVs to mitigate the impact of the 3 well-known types of LA attacks. The controller synthesis is based on its Linear Matrix Inequality (LMI) formulation and the state-space representation of the grid. The contributions of this paper are:
- Developing an EV-based control scheme to mitigate the impact of the 3 known types of LA attacks based on a feedback H-∞ EV-controller formulation.
- Illustrating the effectiveness of the proposed mitigation strategy through time-domain simulations.

The rest of the paper is organized as follows. Section II briefly presents the system preliminaries and the related works. Section III discusses the grid model and the mathematical formulation of the EV-based controller. Finally, Section IV discusses the case studies, and Section V concludes the paper.

## II. Background

In this section, we briefly present the EV numbers and the EV technology needed in our mitigation scheme. We then examine the need for such a mitigation scheme by presenting the three types of LA attacks and a few related studies.

### A. EV Numbers and Technology

The world's governments have been encouraging the adoption of EVs to reduce the transportation sector's emissions to combat climate change. This has led to an exponential growth in EV numbers, demonstrated by the 3.2 and 6.6 million EV sales in 2020 and 2021 respectively [9]. To support this rapid deployment, EVCS manufacturers are continuously introducing faster and cheaper. Level 2 EVCSs now have an 11kW rate [8] while Level 3 fast EVCSs have rates between 40kW and 360kW [8]. While all Level 3 EVCSs are DC chargers, Level 2 can be AC or DC. DC EVCSs are becoming more common owing to their higher charging rates and ability to support the V2G functionality by using bidirectional converters [8].

These EVCSs are connected to the internet using onboard 5G routers. 5G networks are intended to function in areas of high user density and achieve a speed of 10GBps and guarantee a maximum latency of 1ms [10] guaranteeing a reliable internet connection for the EVCSs. Moreover, the Open Charge Point Protocol (OCPP), which is the main standard for the EV ecosystem, species the NTP protocol as the main protocol used to ensure synchronization of the EVCS clocks [11] [12]. NTP ensures the synchronization of the clocks of the EVCSs with a guaranteed accuracy of 10ms over the public Internet [13] and 1ms over the same local area network [13].

*B. LA Attack Types and Attacker Model*

As mentioned above, LA attacks manipulate actual power consumption to cause harm to the grid [4]. The work in [4] proposed a group of static attacks against the grid using a botnet of compromised high-wattage IoT devices. The presented attacks only required high-level geographic information of the devices' distribution. By compromising a small portion of the available smart water heaters, the attackers launched their static attack, sudden spike in load, and caused frequency fluctuations, and line tripping. Moreover, the authors of [5] presented a multi-step static attack by launching a series of static attacks based on the grid's transient condition at each step. This multi-step attack requires the attacker to monitor the grid's transient response to launch the attack steps accordingly.

Switching attacks are another form of LA attacks that can excite certain unstable modes in the power grid such as inter-area oscillation [6]. No topology information is required for this attack. However, during the reconnaissance phase, the attacker needs to determine the frequency of an unstable mode existing in the power grid by introducing a small chirp signal and monitoring the grid's response to calculate the impulse response of the system. The loads are then switched on and off at that specific frequency to excite the unstable mode.

The third form of LA attack is the dynamic attack [7]. During the reconnaissance phase, the attacker gathers information about the grid parameters to build the state-space model of the system. This state-space is then used to craft an attack as a feedback controller that shifts the grid's eigenvalues to unstable operation regions by manipulating the magnitude and oscillation frequency of the compromised load. During the attack in [7], the generators started oscillating against each other and the frequency deviation exceeded the 2.5% deviation limit tripping the generators. The authors of [7] also demonstrated that the success of the attack does not require using 100% accurate grid parameters. However, using inaccurate parameters slightly reduces the attack's impact.

These examples stress the necessity for a protection mechanism to address these new attack vectors. This is especially true since the world is witnessing an increasing ability of attackers to manipulate high-wattage IoT devices as demonstrated by studies performed in partnership with multiple utilities [5] and cyber security companies [14].

*C. Related Studies*

Most grid protection techniques target very specific attack types while disregarding others. The authors of [5], for instance,

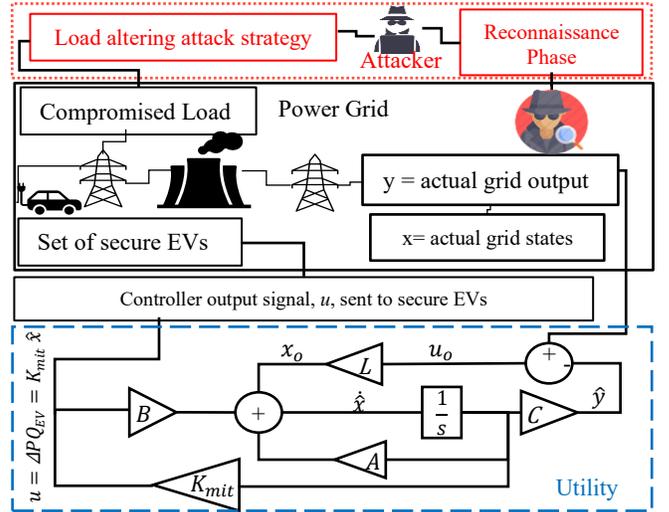

Fig. 1. Defender's state-space model and interaction with the attacked grid

consider that the current N-1 contingency criterion is enough to mitigate the impact of static attacks but proposed a variation of such attacks able to circumvent it and disregard other attack types. The wide area controller proposed in [15] is designed to mitigate switching attacks below 2 Hz making this scheme less effective against higher frequency attacks as well as dynamic attacks. An optimal output feedback controller was used in [16] against inter-area oscillation caused by an individual event on the grid. This work, however, is not meant to mitigate attacks.

Multiple works have considered supporting grid operation by using EVs. The work in [17] for example discusses EV charging at non-unity power factor to inject/draw reactive power and includes it in the optimization of EV charging scheduling [17]. This strategy was able to reduce the overall cost of integration of EVs into the distribution grid and improve voltage stability [17]. Similarly, in [18] bidirectional EV charging is optimized to perform peak shaving for the grid during peak demand times.

III. GRID MODELING AND MITIGATION METHODOLOGY

In the following section, we discuss the defender model and the controller design to be used in our mitigation scheme against LA attacks targeting power grids. The first step towards achieving our mitigation goals would be the representation of the power grid in state-space form. Equally important is the utility's ability to recover the power grid's states by observing its behavior. After establishing the grid representation and the observer design, we introduce our EV-based control synthesis.

*A. Power System Representation and Defender Model*

As a utility, the defender is assumed to have perfect knowledge of the power grid's parameters and is thus able to represent its behavior with extremely high accuracy. As the power system's dynamic behavior is mostly dependent on the generators and their control systems, we use the models that are widely accepted in similar studies [19], i.e., (i) round rotor synchronous machine with order 6, (ii) generator exciter Model IEEE T1, (iii) single mass IEEE G2 steam turbine prime mover, and (iv) Power System Stabilizer (PSS) based on IEEE Std 421.2. Finally, the grid is linearized into a state-space model with the active and reactive power of the aggregate EV-defender loads as the inputs and the generator frequencies as outputs. Additionally, the unknown LA attack is represented by a disturbance to the grid. This representation is expressed in (1) where $x$, $y$, $u$, and $\omega$ represent the vectors of system states, outputs, inputs, and disturbances, respectively. Additionally, $\Delta P_{EV_n}$, $\Delta Q_{EV_n}$, and $f_{Gen_m}$ represent the change in active and reactive power of the aggregate EV load at bus n and the frequency of generator m respectively.

$$\dot{x} = Ax + Bu + B_d\omega \quad (1a)$$
$$y = Cx + Du + D_d\omega \quad (1b)$$
such that $\quad y = (f_{Gen_1} \quad f_{Gen_2} \quad \cdots \quad f_{Gen_m}) \quad (1c)$
and $\quad u = \Delta PQ_{EV} = (\Delta P_{EV_1} \Delta Q_{EV_1} \cdots \Delta P_{EV_n} \Delta Q_{EV_n}) \quad (1d)$

Fig. 1 represents the LA attack against the power grid and the utility's interaction with the grid to attenuate the attack impact. The attacker in the dotted box would use one of the attack models discussed above to launch their LA attacks. The state-space representation in the dashed box is constructed by the utility, based on the power grid parameters, and (1). This state-space representation is then used to calculate the EV gain matrix $K_{mit}$, used to determine the required EV active and reactive power, $u = \Delta PQ_{EV} = K_{mit}x$, for the system to eliminate the impact of the attack/disturbance. The relation between the disturbance and the generator frequency can be expressed as $y = T\omega$, where $T$ is a transfer function written in terms of the state-space matrices. The behavior of $u = \Delta PQ_{EV}$, is then captured and replicated on the set of secure EVs located throughout the grid. The state-space matrices are also used to calculate the observer gain matrix $L$ needed by the utility to recover the grid states, $x$. Since not all the states are measurable, the utility obtains an estimate, $\hat{x}$, by multiplying the grid's outputs, y, by the observer gain $L$. The power of the aggregate EV load is then calculated as $\Delta PQ_{EV} = K_{mit}\hat{x}$. This design changes $A$ to its closed-loop form, $A_{cl} = A + BK_{mit}$. The eigenvalues of $A_{cl}$ define the grid's stability, thus $K_{mit}$ should guarantee acceptable attenuation levels under different attacks.

### B. Observer Design

Since our EV-based LA attack mitigation scheme is based on full-state feedback control, the utility needs to choose an appropriate design for the observer gain matrix, $L$ [20]. With the introduction of $A_{cl}$ and the observer $L$, (1) become (2).

$$\dot{\hat{x}} = A\hat{x} + BK_{mit}\hat{x} + L(y - \hat{y}) + B_d\omega \quad (2)$$

Given the importance of the observer, we choose the Linear Quadratic Controller (LQR) technique [20] for our formulation. LQR is an optimization problem that balances the energy of states vs the control signals. This balance is controlled by the weights Q and R of the states and inputs respectively. The LQR observer is optimized to minimize the cost function J in (3):

$$J = \int_0^\infty x_o^T Q x_o + u_o^T R u_o \, dt \quad (3)$$

where $x_o = (y - \hat{y})$ and $u_o$ is the input being fed into the state-space given that $Q = Q^T \geq 0$ and $R = R^T > 0$. This problem is solved by finding the value of S satisfying the Algebraic Riccati Equation (4) where $A_o = A_{cl}^T$, $B_o = C_{cl}^T$ and $S = S^T \geq 0$.

$$0 = SA_o + A_o^T S - SB_o R^{-1} B_o^T S + Q \quad (4)$$

Equation (4) is quadratic in S and has no trivial solution, but it has a single positive definite solution S that makes the observer stable. Thus, the observer gain $L$ is determined as

$$L = -R^{-1}B_o^T S. \quad (5)$$

### C. Choice of Controller for EV Mitigation Scheme

Unlike previous efforts to use EVs to support the grid in its steady state [17] [18], we use the EVs to support the grid against persisting unknown LA attacks. We create a mitigation scheme that attenuates the impact of the three types of LA attacks by using EVs in a wide area controller. Thus, we synthesize an H-∞ controller as a convex optimization with LMI constraints [21]. H-∞ control minimizes a disturbance's maximum singular value while optimizing the input and output signals' energy. Moreover, by formulating our scheme as a feedback controller, we eliminate the need for an attack detection tool since the controller reacts to the actual grid's states.

### D. H-∞ Controller Design

In this subsection, the EV-based mitigation control scheme is developed, and its mathematical formulation is obtained based on the LMIs of the desired control law [21]. Writing our controller equations as LMIs gives us the flexibility to fine-tune their design and implement complex control schemes. In our study, we utilize the LMI problem in (6) to (10) to calculate the value of the defender gain $K_{mit}$ used by the utility to maintain grid stability by attenuating the impact of the unknown attack. Our control strategy simultaneously minimizes the attack impact and the EV load involved in the feedback signal. We first write the transfer function $T$, representing the influence of disturbance $\omega$ on the grid output $y$, generator frequency, as (6):

$$T = \frac{y}{\omega} = (C + DK_{mit})(sI - (A + BK_{mit}))^{-1}B_d + D_d \quad (6)$$

The H-∞ controller aims to minimize the maximum singular value of $T$. This translates to minimizing the value of $T$ less than a scalar $\rho$ corresponding to the attenuation level. Based on Schur's formulation for partitioned matrices [21], the problem of minimizing $T < \rho$ is arranged as the LMI problem (7)-(9).

$$\text{Minimize } \rho \quad (7)$$
$$\text{s.t.} \quad X > 0 \quad (8)$$
$$\begin{bmatrix} (AX + BW)^T + AX + BW & B_d & (CX + DW)^T \\ B_d^T & -\rho I & D_d^T \\ CX + DW & D_d & -\rho I \end{bmatrix} < 0 \quad (9)$$

This LMI optimization problem can only have a solution if there exists a matrix W and a symmetric positive definite matrix X satisfying the matrix inequality problem in (7)-(9). After solving for $W$ and $X$, we calculate the feedback gain matrix as $K_{mit} = WX^{-1}$. To guarantee the performance of the controller in the time domain and reduce the settling time, we add constraint (10) based on the D-stabilization pole placement method [7] to select the region of the closed loop eigenvalues.

$$AX + XA^T + BW + W^T B^T + 2\Lambda_1 X < 0 \quad (10)$$

Constraint (10) shifts the eigenvalues of the system into a region where the real part of the eigenvalues is less than $-\Lambda_1$ to guarantee performance and fast settling time. $\Lambda_1$ should also be small enough to avoid aggressive controller behavior.

### E. EV User Impact and Privacy Concerns

This scheme does not require a change in user behavior. The utility will only utilize EVs that are connected to EVCSs at the instance of attack. Furthermore, an 11kW EVCS delivers a charge of 0.9 miles/min. Thus, mitigating an attack lasting a few seconds will have an unnoticeable impact on the EV's ranges. Such a scheme should only involve EVs whose owners gave consent to participate through a utility-managed incentive program. Hydro-Quebec implements a similar program, the Hilo project [22], which allows them to control home loads to reduce demand during peak times with plans to extend it to EVs.

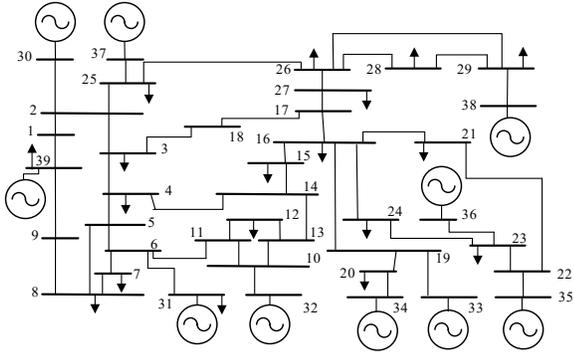

Figure 4. New England 39-bus grid

## IV. CASE STUDIES AND SIMULATIONS

This case study evaluates the performance of our EV-based mitigation scheme against LA attacks on the New England (NE) 39-bus grid [23] presented in Fig. 4. The NE grid has 39 buses, 10 generators, and 19 loads with a total of 6,097MW. The simulations were performed using MATLAB-Simulink Specialized Power Systems Toolbox with $1 \times 10^{-9}s$ step size.

While we acknowledge that the current number of EVs is not enough to exploit the mitigation scheme's full potential, the following example demonstrates its future feasibility. Thus, we choose the similarly sized New South Wales (NSW) grid [24] having a 6968 MW load [24] and 5,892,206 registered vehicles [25]. Scaled to fit our test grid, the number of vehicles is 5,155,681. At a future 50% penetration level, our grid will have 2,577,841 EVs. As per the International Energy Agency (IEA) [9] there is 1 public EVCS for every 10 EVs on the road. This means that there will be 257,785 public EVCSs in our grid.

With an average EVCS occupancy rate of 31% [26], we can estimate that there will be 80,000 EVs connected to the grid on average. Furthermore, based on the mixture of different EVCSs, the average charging rate per EVCS is 24kW [9]. Based on these statistics, we estimate there will be 1,920MW of EV load connected to the grid. Furthermore, there are roughly 9 private EVCSs for every 10 EVs [9], which means 2.3 million private EVCSs in our grid. This demonstrates that the total EV load can easily surpass the needed levels for the success of our proposed EV-based LA attack mitigation scheme. Additionally, with the increased adoption of DC EVCSs, the presence of V2G capabilities will only increase in the future. However, in the absence of V2G, our scheme still holds, but requires double the number of EVs. Instead of starting V2G, the controller sends a command to another set of EVs to stop charging.

### A. Reduced Power Grid State-Space Model

Due to the complexity of the grid, the state-space matrices size is large making controller design prohibitive, i.e., our grid has over 300 states. To preserve correct system behavior while reducing its order, we use Hankel model reduction [27] on the state-space matrices. Hence, we calculated the Hankel Singular Values of the states and decided to reduce the system order to 30, preserving 99.81% of its energy. A further increase in order does not significantly improve accuracy, while significantly increasing design complexity. This reduction is only used for controller synthesis. However, the simulations are performed on the actual grid, not the linearized or reduced model.

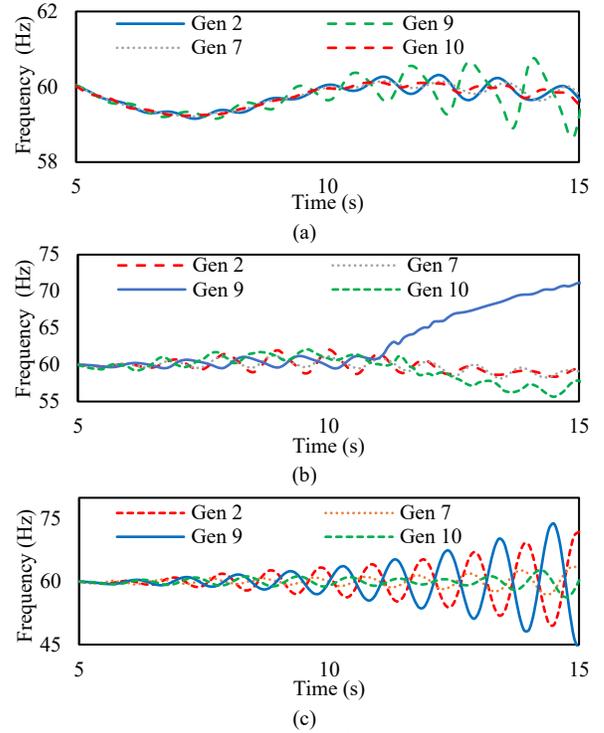

Figure 2. Generator frequency response with no PSS under (a) Attack 1, (b) Attack 2, and (c) Attack 3

### B. Case Study 1: Attacks in Absence of Mitigation Scheme

Based on the work presented in [4], [6], and [7], which discuss static attacks, switching attacks, and dynamic attacks respectively, we formulate our attack scenarios against the power grid in the absence of the suggested mitigation strategy to demonstrate their impact as a base case. The detailed mathematical formulation of the attacks is found in [4], [6], and [7]. The first batch of attacks is simulated in the absence of the PSS and the impacts are discussed in the following three cases.

1) Attack 1: is an 800MW static attack initiated at t=5s split on buses 3, 4, 24, and 29. This attack is a single spike in load equal to 13% of the system demand. The attack causes a drop in frequency and the oscillations presented in Fig. 2(a).

2) Attack 2: is an 800MW switching attack split equally on buses 3, 4, 24, and 29. Fig. 2(b) demonstrates how this attack caused the generators to oscillate against each other and a large frequency deviation that reached 71Hz. This attack would cause the generators to trip resulting in a blackout.

3) Attack 3: is an 800MW dynamic attack split on buses 3, 4, 24, and 29. In this attack, the loads do not always oscillate in phase. Fig. 2(c) demonstrates the attack impact in which the generators' frequencies oscillate wildly and continue to grow to reach 74Hz. This trip the generators and cause severe damage.

The second batch of attacks repeats the same three scenarios but in the presence of the PSS. The PSS is meant to stabilize the grid by damping generator frequency swings, but as we demonstrate, it falls short of eliminating LA attack impacts.

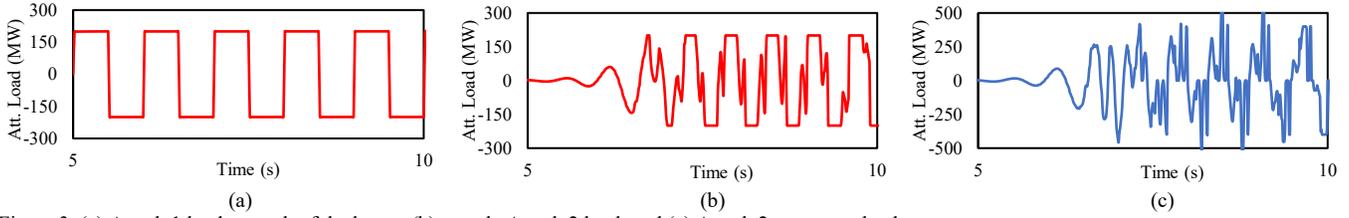
Figure 3. (a) Attack 1 load on each of the buses, (b) sample Attack 2 load, and (c) Attack 2 aggregate load

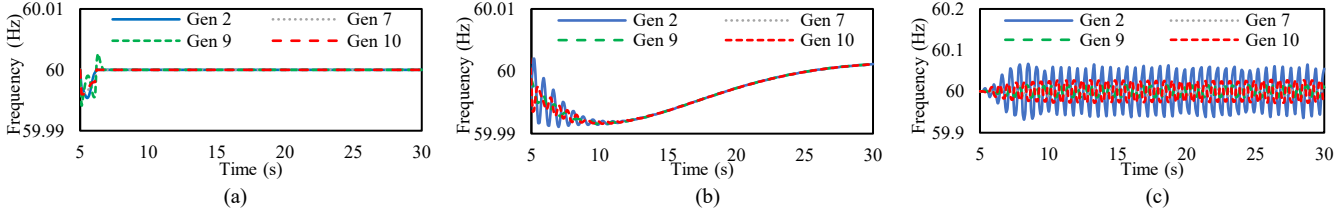
*Figure 5. Generator frequency response in the presence of the proposed H-∞ controller under (Left) Attack1, (Center) Attack 2, and (Right) Attack 3*

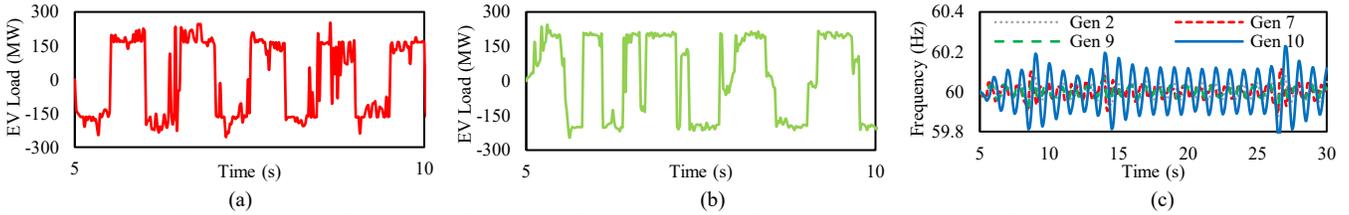
Figure 6. Aggregate EV mitigation load for (a) Attack 2, (b) Attack 3, and (b) generator frequency under attack 2 with 25% less defender load

1) Attack 1: is repeated in the presence of the PSS and causes a frequency drop and minor oscillations that are eliminated by the actions of the PSS and the system regains its stability.

2) Attack 2: is repeated in the presence of the PSS yet the system sustained oscillations reaching a deviation of 1 Hz. Such an attack causes the operator to shed 5% of the total grid's load and trip the generators if sustained for 3min. Fig. 3(a) demonstrates a sample Attack 2 load on bus 4.

3) Attack 3: is repeated as well in the presence of the PSS and causes rapid frequency oscillation and a large deviation of 1.5 Hz. This attack will instantaneously trip the generators leading to a blackout. Fig. 3(b) demonstrates a sample Attack 3 load and Fig. 3(c) represents the aggregate of the 4 attack loads.

This case study demonstrates that even in the presence of the PSS, the LA attack impact is eliminated when the attack is crafted properly. Thus, adding a mitigation mechanism that eliminates the impact of such attacks becomes necessary.

### C. Case Study 2: H-∞ Control EV-Based Mitigation Scheme

Based on our methodology, we design our EV mitigation scheme and test it against the three attacks. The EV defender load is calculated based on $\Delta PQ_{EV} = u = K_{mit}\hat{x}$ and replicated on the NE grid that is under attack. The EV mitigation scheme successfully eliminates the static attack's impact immediately after its initiation even if it is sustained as demonstrated in Fig. 5(a). The switching and dynamic attacks are presented below.

1) Attack 2: is repeated against the NE 39-bus grid in the presence of the H-∞ EV-based control scheme. The frequency responses of the generators are presented in Fig. 5(b). The frequency drops by 0.01 Hz and recovers to 60 Hz at t=30s, representing 100% system recovery. Fig. 6(a) presents a sample of the aggregate EV mitigation load on bus 4.

2) Attack 3: is repeated in the presence of the H-∞ EV-based control scheme. The EV controller severely reduces the attack impact as demonstrated by the frequency response presented in Fig. 5(c). The frequency deviation was reduced to 0.08Hz representing a 94.7% reduction of the attack impact. Fig. 6(b) presents a sample of the aggregate EV mitigation load on bus 4.

In both cases, the frequency returns to the normal operating range. For the switching attack, the frequency returns to 60Hz, but for the dynamic attack, the frequency keeps oscillating in a $60 \pm 0.08$Hz range which is considered a safe operation range that does not require engaging the protection mechanisms.

We repeat the attacks when the defender does not control any EVs on the attacked bus. This is to demonstrate the success of the mitigation scheme even if we eliminate one of its advantages, i.e., the colocation of the defender-controlled EVs with the attack load. Attack 2 and Attack 3 are both repeated against the grid that has the EV mitigation scheme and the frequency deviation reached 0.08Hz and 0.09Hz respectively.

The impact of the synchronization error is also studied on the mitigation scheme by repeating Attack 2 after adding a random Gaussian delay to each of the feedback signals between 0ms and 10ms. The maximum frequency deviation that was observed in this case was 0.07 Hz. This shows how the mitigation scheme is still successful even in the presence of the maximum delay possible by using NTP protocol. To quantify things in terms of controller success, the maximum frequency deviation was reduced by 91% when compared to the case with a PSS but in the absence of our proposed mitigation scheme.

### D. Smaller Defender Load

We now examine the performance of our mitigation scheme when the utility has fewer EV resources than the attack load. Similar studies agree that the utility always has larger resources in its own grid than attackers. However, we consider a scenario in which attackers successfully compromise enough EVs such that the EV load controlled by the utility is smaller than the attack load. This requires attackers with huge resources. In this case, the defender partially mitigates the attack impact. Fig. 6(c) demonstrates the impact of Attack 2 when the utility controls a 25% smaller EV load than the total attack load. This attack causes a 0.23Hz frequency deviation, which is significantly smaller than that in the absence of our EV-based mitigation and does not trigger any generator tripping or load shedding.

### E. Impact on EV Range

We now evaluate the impact of our mitigation scheme on the EVs' range. This evaluation is based on the average charging rate of 24kW. Mitigating Attacks 2 or 3 requires the EV to alternate between charging and V2G such that the net charge is approximately 0. Attacks 2 and 3 cause a loss of 0.001kWh and 0.009kWh respectively. The EVs also lose the opportunity to charge 0.2kWh. The total loss is equivalent to a range of 1 mile for each EV. For an EV that was connected to an EVCS but not charging, the net impact is almost 0kWh. Attack 1 on the other hand will result in a total loss of 0.4kWh or 2 miles.

### F. Mitigation Scheme Feasibility

An added advantage of using EV charging in our mitigation scheme is that the required communication and control infrastructure is already in place. The Central Management System (CMS) that already exists in the EV ecosystem communicates and controls all the public EVCSs in real-time. Using the OCPP protocol, the CMS can turn the individual EVCSs on, off, change their charging rate, and discharge using V2G [11] [12]. This means that adopting our mitigation scheme would not require the addition of any software, hardware, or communication to the ecosystem. The Hilo project gives the utility the same capabilities over private EVCSs as well.

Mitigating these attacks using our suggested scheme requires an extremely minimal cost to be incurred by the utility. To demonstrate this, we consider the average 24kW EVCS in Quebec, Canada [28]. The hourly price of charging on these EVCSs is 7.53 CAD billed per second [28]. This means that mitigating Attack 2 and Attack 3 costs the individual EV user 6.28 cents. This means that when the utility reimburses the EV users for this cost, the utility will have to pay a total of 2072 CAD (1,530 USD) plus whatever extra value the utility determines in its incentive program. When compared to the millions of dollars of losses incurred during blackouts, the price of our suggested scheme is virtually insignificant.

## V. Conclusion

In this paper, we presented the advantage of using EVs as a feedback controller used to mitigate the impact of the three types of LA attacks. We performed the study from the perspective of the utility that has an accurate state-space model of its grid. The utility uses this model to design the EV feedback gain based on an H-∞ controller formulation. The proposed mitigation scheme was tested against the three classes of LA attacks and achieved grid stability against all three classes. The EV-based controller lowered the frequency deviation caused by 800 MW LA attacks below 0.1Hz. The behavior of the controller was then studied under different operating conditions that lowered its effectiveness but still achieved grid stability.

## Acknowledgment

The work of Ph.D. candidate M. A. Sayed is supported by Fonds de Recherche du Quebec-FRQNT, project 2022-2023 - B2X - 317973. This research was conducted and funded as part of the Concordia University/ Hydro-Quebec/ NSERC research collaboration project Grant reference: ALLRP 567144-21.